# Excitations of Incoherent Spin-Waves due to Spin-Transfer Torque

K. J. Lee[1, 2*], A. Deac[1,2], O. Redon[1,2], J. P. Nozières[1] and B. Dieny[1]

[1]*SPINTEC, CEA/Grenoble, DRFMC, 38054 Grenoble, France,*

[2]*CEA/Léti, 38054 Grenoble, France*

* Permanent Address: Storage Lab., Samsung Advanced Institute of Technology, Suwon, Korea

**As predicted by Slonczewski[1] and Berger[2], the possibility of exciting microwave oscillations in a nanomagnet by a spin-polarized current has been recently demonstrated[3]. This observation opens very important perspectives of applications in RF components. However, some unresolved inconsistencies are found when interpreting the magnetization dynamics within the coherent spin-torque model (CSM)[4-6]. In some cases, the telegraph noise caused by spin-currents could not be quantitatively described by CSM. This led to controversies about the need of an effective magnetic temperature model (ETM)[7-11]. Here we interpret the experimental results of ref. 3 using micromagnetic simulations. We point out the key role played by incoherent spin-waves excitation due to spin-transfer torque (STT). The incoherence is caused by spatial inhomogeneities of local fields generating distributions of local precession frequencies. We observe telegraph noise with GHz frequencies at zero temperature. It is a consequence of the chaotic dynamics and is associated with transitions between attraction wells in phase space.**

In the interpretation of the experimental results of ref. 3 within CSM, three puzzling results are obtained: i) the existence of a region in the experimental phase diagram which cannot be explained within CSM (labelled W in ref. 3), ii) a rather low $4\pi M_{eff}$ (= 0.68 T) instead of the well-known value for Co (~ 1.5 T) required to fit the small amplitude ferromagnetic resonance (FMR) frequency (We ourselves experimentally checked on similar samples that such a reduction was largely overestimated even by including an interfacial out-of-plane anisotropy), iii) broad spectra with multiple peaks (for example, point 5 of Fig. 2d of ref. 3). Full micromagnetic model (FMM) including STT term has already indicated that non-uniform dynamics of magnetization[12] and chaotic behaviour[13] can be caused by STT. Here we show that the spatiotemporal incoherence in the magnetic dynamics induced by STT can naturally explain the experimental results of ref. 3.

Fig. 1a shows a calculated contour of normalized magnetoresistance, $\Delta R/R_{max}$ = $(R_{ave}-R_{parallel})/(R_{antiparallel}-R_{parallel})$, where $R_{ave}$ is proportional to the average magnetization component along the long axis of the sample (elliptical shape, ~ 130nm by 70nm). In the precession region, $R_{ave}$ is obtained from the time-average of the resistance variations over 100ns. Fig. 1a shows a good semi-quantitative agreement with Fig. 2a of ref. 3 along the current axis. In particular, it reproduces the "mysterious" region labelled "W" of the stability phase diagram. However, the agreement between our calculations and the experiments is not perfect in the position of switching-precession boundary (~Hc, coercivity). The difference may originate from underestimated anisotropy fields in the simulation, and possible deviations of the sample size from the nominal one in experiments. FMM reveals that the "W" region corresponds to dynamic vortex formation/annihilation due to the interplay of the large current-induced field and spin torque. Fig. 1b shows the very large temporal variation of the normalized modulus of the magnetic moment of the sample. The moment remains close to zero when the vortex forms in the intervals labelled "V" and from time to time increases close to unity when the vortex annihilates.

In the switching region, the single domain model (SDM) shows that STT excites coherent precession modes which eventually lead to magnetization switching (Fig. 2a). However, a more complex dynamics is observed in FMM. Before switching, three consecutive stages exhibiting quite different magnetization motions can be distinguished: i) growth of the precessing end domains (stage I), ii) steady precession of the end domains (stage II), and iii) chaotic domain motion (stage III). Just after turning on the current, the magnetizations at the two long edges of the cell start precessing (stage I, Fig. 2b and 2c). Incoherent spin-waves are first excited at the edges of the long axis because of spatially non-uniform local demagnetizing fields. The difference in the local fields between the cell centre and the edges is of about 3 kOe. The precessing end domains become broader and broader versus time. At this stage, the resistance oscillations are a bit asymmetric because of clockwise current-induced fields. Once the end domains have almost joined each other, they keep on precessing for a while (stage II, Fig. 2d and 2e). The magnetizations at the cell centre still remain along the initial direction, because the local torque from either magnetic field or spin-current is still too weak. At the beginning of stage III, the magnetizations at the cell centre start precessing. Then, a growing spatial incoherency in precession frequency and magnetization orientation is observed. Consequently, the domain motion becomes chaotic (stage III, Fig. 2f and 2g), and finally the magnetization switches.

Fig. 3a shows a contour of the microwave power divided by square of the current ($I^2$) obtained in FMM. Interestingly, we could almost duplicate the small angle FMR frequencies of ref. 3 (Fig. 3b). Thus, by properly taking into account the incoherent spin-waves, the experimental FMR frequencies can be reproduced without having to use an artificially reduced value of 4πMs for Co in Kittel's formula[14] established within SDM. FMM shows that a coherent precession of the magnetization is observed in a narrow range of current which weakly depends on the magnetic field between 0.3 and 2kOe. This current range is limited on the lower side by Ic~1.6mA which characterizes the onset of small angle magnetization precession and on the upper side by $I_T$~2.0mA associated with the onset of the incoherence in the excitations (Fig. 3c). It broadens with decreasing size of the nanomagnet and/or ratio of magnetostatic to exchange energy since the incoherence is a consequence of the competition between magnetostatic and exchange energy. If we now consider the large amplitude excitations, in SDM, large amplitude precession modes are expected in a broad range of currents and fields.[5] In FMM, however, a relatively large amplitude dynamics is obtained within a limited range of current and field (red and yellow region of Fig.3a). The microwave spectrum at point A (I=3 mA) on line 2 in Fig. 3a corresponds to a small angle precession with an average angle of 6 degree. Besides a relatively broad peak centred around 13 GHz, it exhibits a large low frequency response like 1/f noise due to the chaotic nature of the incoherent spin-waves excitation as mentioned above (see Fig.3d). At point B (I=7 mA) where the maximum power is obtained, the spectrum exhibits the largest peak at a relatively low frequency (about 1 GHz). Note that the spectrum at point B is multiplied by a factor 0.2. However, compared with the spectrum derived from SDM (inset of Fig.3d), the amplitude and width of the largest peak is much smaller and broader. The integrated power over 0.1-30 GHz in FMM is only 58% of that in SDM. Further increase in the current introduces even more incoherence in the spin-waves excitation.

Transient unstable vortex configurations are observed at high current densities. As a result, the spectrum at large current shows only 1/f-like noise without any peaks in the high frequency range (point C (I=14 mA), Fig.3d). The simulation results along line 2 almost duplicate the reported experimental results (Fig. 2d of ref. 3). In the higher field regime (not shown), we obtained similar trends in the shape of the spectra such as the 1/f-like noise and small/broad peak structures at high frequency range.

An interesting feature in the spectrum at point B is that the largest peak is obtained at 1 GHz. Such low frequency peak is rather surprising because the actual precession frequency of local magnetization is of about 10 GHz (see a smaller and broader peak in the spectrum around 10 GHz in Fig. 3d, curve B). As shown in Fig. 4a, the unexpected low-frequency dynamics corresponds to random fluctuations between almost parallel and antiparallel magnetic configurations. In the random fluctuation patterns, the dynamics of the low resistance state (indicated by A) is different from that of the high resistance state (indicated by B). It is because the instabilities in the former and latter states are respectively driven by the STT (spin-waves) and the external field (no spin-wave). It should be noted that the telegraph noise was obtained using zero temperature calculations. CSM[6] cannot describe telegraph noise at zero temperature since the dwell time in a state is proportional to $\exp[U(I,H)/k_BT]$ where $U(I,H)$ is the modified energy barrier due to STT. Therefore, one may naively think that the chaotic dynamics is somewhat equivalent to an effective magnetic temperature (ETM) ($T_m$). Following the concept of ETM, we tried to estimate $T_m$ from the reduction in the normalized modulus of the nanomagnet total magnetic moment $|M|/M_s$ using equation (1)[15],

$$\frac{|M|}{Ms} = 1 - \frac{1}{4\pi^2 S}\left(\frac{k_B T_m}{2JS}\right)^{3/2} \frac{\sqrt{\pi}}{2} \zeta\left(\frac{3}{2}\right) \qquad (1)$$

where the spin-wave stiffness constant $2JSA^2$ is of the order of 3 meVnm$^2$ in our simulation, $A$ is the cubic lattice constant of Co, and the zeta function $\zeta(3/2)$=2.61. We found that $T_m$ is negligible at I < $I_T$, but abruptly increases with current at I > $I_T$ (Fig. 4b). The dependence of $T_m$ versus I is found to be independent of the applied field over the large range of field investigated, indicating that the current solely determines $T_m$. The slope is of about 500 K/mA at current slightly larger than $I_T$, and becomes smaller at higher current. These values agree well with the experimentally reported values of $T_m$[7,9]. However, to further evaluate the validity of this concept of the effective magnetic temperature, we compared the spin-waves spectra generated at T=0K by STT (I=5mA, H=600Oe) corresponding to an effective magnetic temperature of 800K with the thermally excited spectra obtained at an actual temperature of 800K (I=0mA, H=600Oe). The spectra were calculated by time-averaged spatial Fourier transformations of magnetization (= <$m_q$> in Fig. 4c). As shown in Fig.4c, significant differences in the spin-wave spectra induced by STT and thermal excitation exist. The STT generates larger (smaller) population of spin-wave modes at low (high) wave-number (q) than the actual thermal excitation. We therefore conclude that the concept of ETM provides a convenient representation of the overall density of excitations generated by STT but does not give a good quantitative description of the spectra of these excitations. As a consequence, it is more accurate to explain the observed telegraph noise by the chaotic dynamics itself rather than by the concept of ETM. The non-linear magnetization dynamics induced by STT and in particular the incoherent excitations, leads to the formation of attraction wells in phase space in ranges of field and current larger than coercive values i.e. where no energy minimum is allowed in CSM (Fig. 4d). These trajectories of higher probability in phase space are known as strange attractors in chaotic dynamics[16]. Random transitions between these attraction wells with about GHz frequencies (which are not necessarily associated with energy minima since no static stable state is allowed in this range of current and field) quantitatively explain the recent experimental observation of telegraph noise at I > Ic and H > Hc[10].

Magnetic dynamics excited by spin current in nanopillars can be chaotic over a wide range of currents and fields leading to a broadening of linewidths or even 1/f-like noise. Recently microwave spectra with sharp peaks were observed[17] in point contact experiments. In these experiments, the physical edges of the samples are far from the excited region due to the point contact geometry. As a result, the effective field is more homogeneous over the excited region so that the magnetic excitations can develop in a more uniform way eventually leading to a more coherent precession of the magnetization.

**Methods**

Slonczewski's expression of the additional term, $\left(\partial_t \vec{M}\right)^{STT} = (\gamma a_J / Ms)\vec{M} \times (\vec{M} \times \hat{p})$ [1], due to STT was adopted in the conventional Landau-Lifshitz-Gilbert (LLG) equation. Here, $\gamma$ is the gyromagnetic ratio, $\vec{M}$ is the magnetization vector of the free layer, $\hat{p}$ is a unit vector parallel to the electron polarization, $Ms$ is the saturation magnetization, and $a_J$ is the amplitude of the spin torque in the unit of magnetic field[4]. In this work, we did not take into account any angular dependency of $a_J$. The following parameters were used for the free layer: The layer has an elliptical shape with long (respectively short) axis of 129.6 nm (respectively 72.0 nm), the thickness is 3 nm, the saturation magnetization $Ms$ is emu/cm$^3$, the exchange stiffness constant is 2 ×10$^{-6}$ erg/cm, the anisotropy field is 30 Oe, the spin polarization is 0.4, the intrinsic damping constant is 0.014, and the unit cell size is 3.6 nm. All parameters were chosen to mimic the experiments of ref. 3. The relatively low value of the exchange stiffness constant for Co was chosen to mimic the room temperature measurements[18]. When we used the usual higher value for hcp Co (= 3.3×10$^{-6}$ erg/cm), only shifts of boundaries in the phase diagram were obtained but the main conclusions of this work were not altered. The other assumptions were: collinear polarization of incoming electrons with the pinned layer magnetization set along the long axis, no stray field from the pinned layer on the free layer; initial parallel configuration of magnetizations; zero temperature; and electrons flowing from the free to the pinned layer. The current-induced magnetic field was included. In-plane external fields were applied along the long axis with a tilt angle of 1 degree. Positive external fields prefer the parallel configuration of magnetizations. For the stochastic calculation, the Gaussian-distributed random fluctuation fields (mean=0, standard deviation = $\sqrt{2\alpha k_B T /(\gamma M_S V \Delta t)}$, where $\Delta t$ is the integration time step, $V$ is the volume of unit cell.)[19] have been added to the effective fields of LLG equation.

To test the convergence, we computed the power spectrum using different grid sizes. When we reduced the grid spacing by a factor of two, no significant changes were observed. Nevertheless, due to the non-linear character of the present dynamics, it was impossible to get exactly the same detailed time-trajectories in phase space with different cell size but the overall features such as the excitations spectra were identical.

On the contrary, if the cell size was increased by a factor 2, the results were significantly different even for the excitations spectra.

**Acknowledgements** We acknowledge A.Vedyaev, U. Ebels, I. N. Krivorotov for very fruitful discussions. This work was supported by the RTB project of CEA, Léti. Correspondence and requests for materials should be addressed to K. J. Lee (lee@drfmc.ceng.cea.fr).

Figure captions

Figure 1: Calculated phase diagram of the normalized magnetoresistance ($\Delta R/R_{max}$) with varying currents and fields. (a) Obtained in Full MicroMagnetic model (FMM), and (b) Dynamic variations of the normalized modulus of the total moment ($=|M|/M_s$) at I = 14 mA and H = 600 Oe. "V" indicates the interval of almost zero moment associated with transient vortex formation.

Figure 2: Magnetization switching due to spin-transfer torque, (a) resistance variations versus time at I = 4 mA and H = 0 Oe. The (b)-(f) show magnetic domain patterns of FMM at each time stage; Stage I: (b) 0.20 ns, (c) 0.30 ns. Stage II: (c) 1.5 ns, and (d) 1.55 ns. Stage III: (e) 3.2 ns, and (f) 4.3 ns.

Figure 3: Contour of microwave power normalized by $P_0$ divided by $I^2$ in the frequency range of 0.1-30 GHz. The plotted power is the average power generated in steady state, and $P_0$ is a constant reference power over the whole ranges of current and field introduced for dimensionless unit in logarithm: (a) FMM, and (b) Comparison of small amplitude FMR frequencies obtained along line 1 of (a) (I = 2 mA). (c) coherent precession range of bias current (H = 400 Oe). The microwave spectra corresponding to points A-C in (a) are shown in (d). Lines 2 corresponds to H = 400 Oe. Inset of (d) shows a microwave spectrum obtained from the single domain model at I=7 mA, H=600 Oe. The curves in **c** are vertically offset.

Figure 4: Telegraph noise at zero temperature and effective magnetic temperature. (a) Variations of the normalized differential resistance versus time at the range of current and field giving the maximum $P/I^2$ in FMM (I=7 mA and H=400 Oe). (b) Effective temperature as a function of current. (c) Time-averaged population of spin-wave modes in magnetic dynamics excited by spin-transfer torque and thermal activation. (d) Time-trajectories in phase space showing strange attractors, obtained at I=7 mA and H=400 Oe.

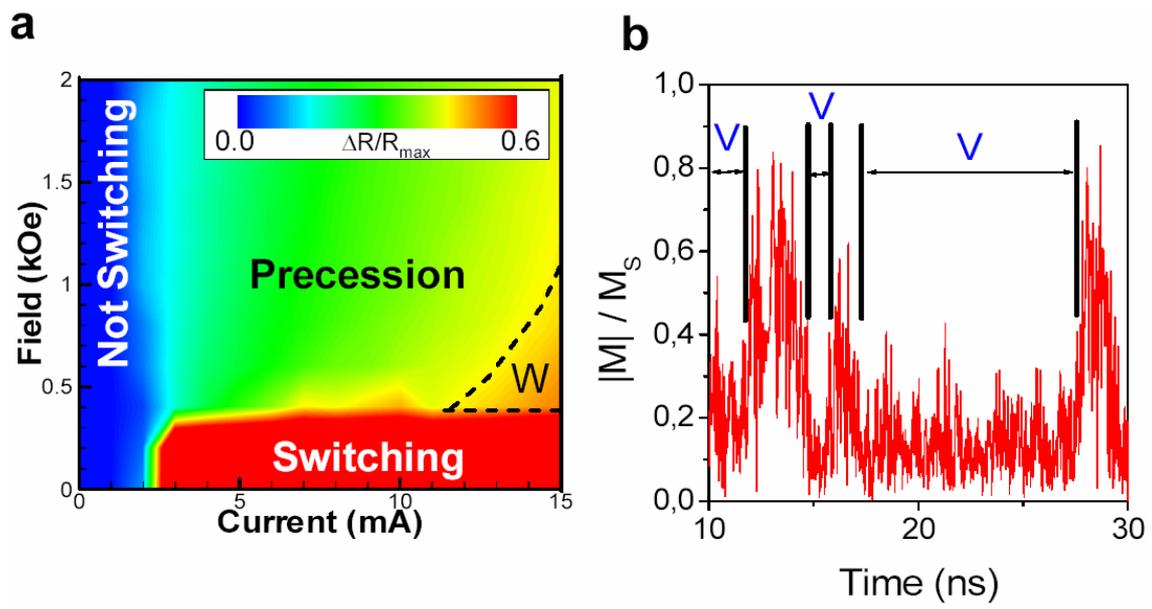

Fig. 1. K. J. LEE et al.

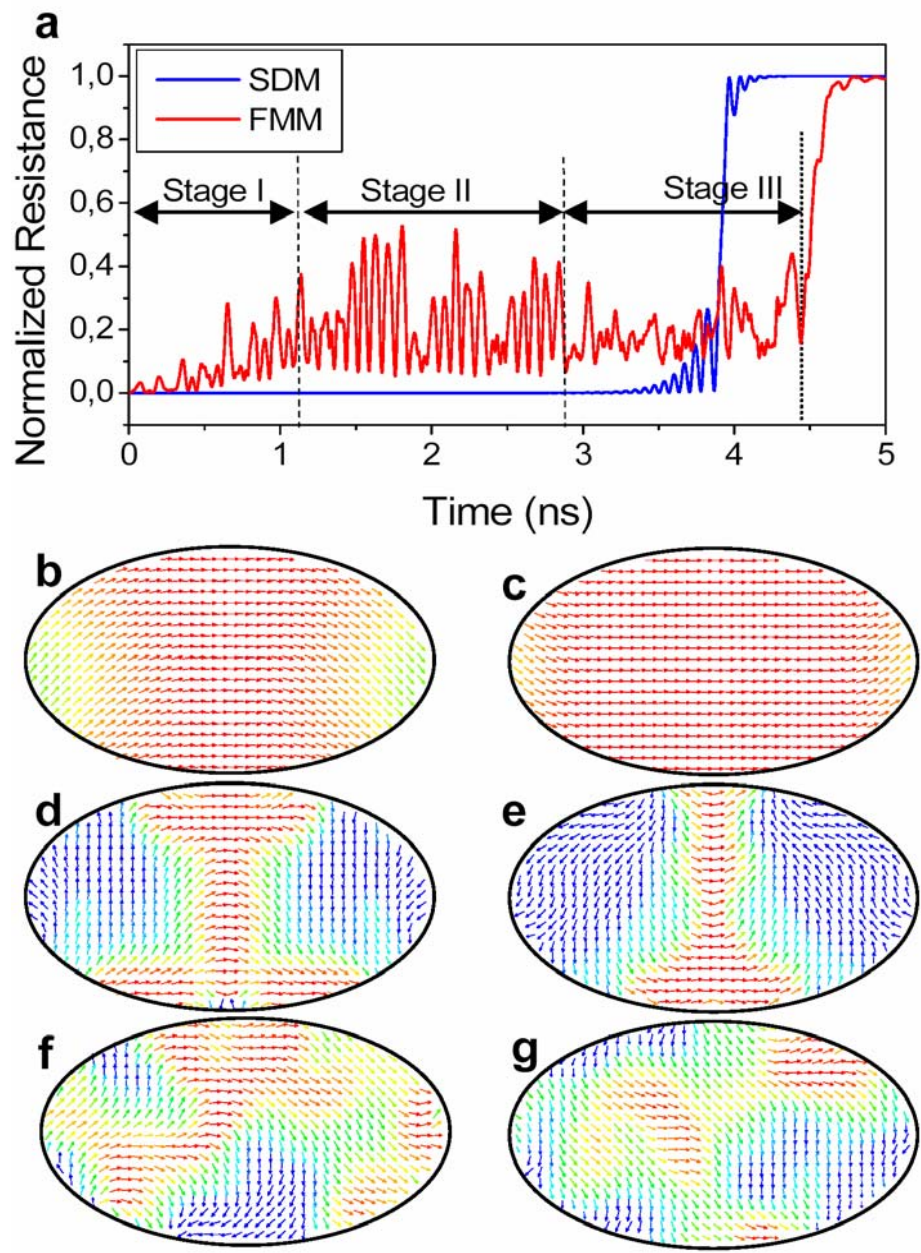

Fig. 2. K. J. LEE et al.

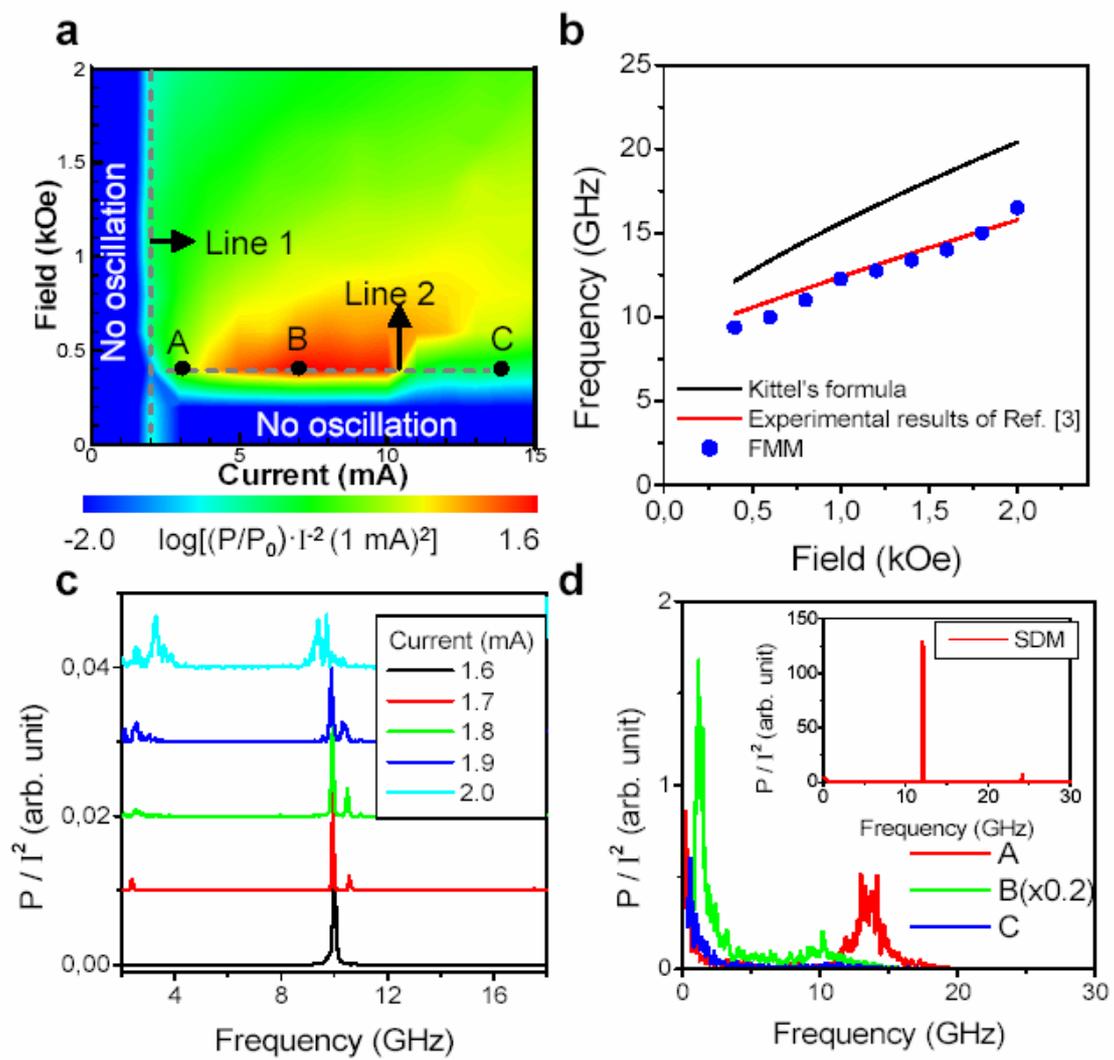

Fig. 3. K. J. LEE et al.

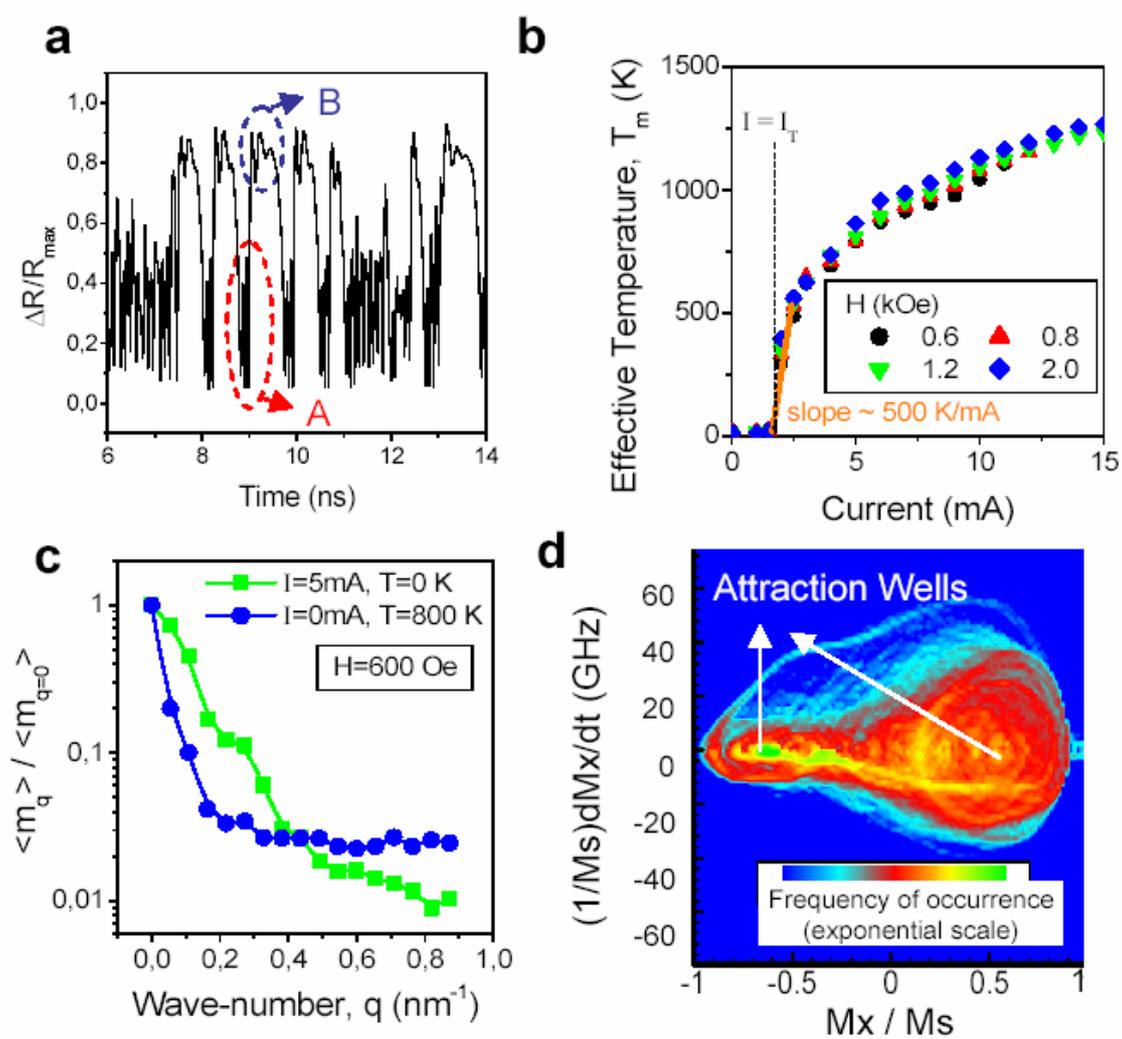

Fig. 4. K. J. LEE et al.